\documentstyle[12pt]{article}
\newcommand{\be}{\begin{equation}}
\newcommand{\ee}{\end{equation}}
\begin{document}
\baselineskip=24pt
{\centerline {\bf ANYONS IN ELECTROMAGNETIC FIELD AND THE BMT EQUATION}}

{\centerline {\bf Subir Ghosh}}
\centerline {Physics Department, Gobardanga Hindu College}
\centerline {24 Pgs. (North), West Bengal}
\centerline {India}
\vspace {1 cm}

\centerline {Abstract}
The Lagrangian model for anyon, presented in [6], is extended to include
interactions with external, homogeneous electromagnetic field. Explicit
electric and magnetic moment terms for the anyon are introduced in the
Lagrangian. The 2+1-dimensional BMT equation as well as the correct value
(2) of the gyromagnetic ratio is rederived, in the Hamiltonian framework.

PACS Numbers: 11.10.Ef
\vfil\eject

During the last decade, 2+1-dimensional physics has been an area of intense
activity. This is mainly due to the application of anyons, (2+1-dimensional
particles of arbitrary spin and statistics), in realistic planar physics [1],
notably fractional quantum Hall effect, high $T_c$ - superconductivity etc.

This has provided the impetus for constructing viable classical models
of free, relativistic, spinning particles in 2+1-dimensions. Starting
with the work of Jackiw and Nair [2], who have derived a covariant equation
of motion for free anyon, (in analogy to the Dirac equation for spin ${1\over
2}$
particle), a number of papers have appeared in this connection [3, 4]. However
[4] does not develop a configuration space Lagrangian framework. This was
attempted
in [3], but unfortunately the work has given rise to criticisms [5]. On the
other
hand, the model proposed by the present author [6], following the authoritative
work of Hanson and Regge [7], adequately describes a free anyon.

But the interesting physical properties of anyons, in particular its
gyromagnetic ratio $g$, can only be probed in the presence of
electromagnetic interactions. This was first done by Chou, Nair and
Polychronakos [8],using the symplectic framework for the induced representation
of the Poincare group for anyon, minimally coupled to external electromagnetic
field. They found $g = 2$ for anyon, by comparing their equations with the
2+1-dimensional Bargmann-Michel-Telegdi (BMT) [9] equation. However they
have not derived an explicit form of the Lagrangian.

In the present Letter, we have developed a {\it Lagrangian} model for
a 2+1-dimensional spinning particle, (anyon), with {\it electric and magnetic
dipole moments},
interacting minimally with external electromagnetic field.
Several
attractive features of the present formulation and the results obtained
are the following: (i) A coordinate space, reparametrization invariant
Lagrangian for the interacting model is provided. Detailed constraint
analysis is performed in the Hamiltonian formulation and the BMT equation
is obtained via Dirac brackets [10]. Albeit classical, the BMT equation
is in fact the same as the quantum mechanical Heisenberg equation of motion
for the spin. We also find $g=2$ for anyons.

(ii) For simplicity, as well as our goal of fixing the $g$ value, we
produce only $O(F_{\mu\nu})$ results, where $F_{\mu\nu}$ is the external,
uniform field. All these restrictions can be lifted in a straightforward
manner. A kinetic term for the gauge sector can also be included to get
the full quantum theory. In fact, as we comment at the end, inclusion
of the gauge kinetic term might be imperative for conceptual reasons, as
has been pointed out in [7].

After this brief outline of the present work, we start with the Lagrangian
$$ L=(M^2 u^2 +{1\over 2}J^2\sigma^2 + MJ\epsilon^{\mu\nu\lambda}u_\mu
\sigma_{\nu\lambda})^{1\over 2} -eu_\mu A^\mu -{D\over 2}\sigma_{\mu\nu}
F^{\mu\nu} $$
\be = ({\cal L})^{1\over 2} -eu.A -{D\over 2}\sigma.F .\ee
where ${\cal L}$ is the same as that of the free anyon in [6] and the variables
are also defined as in [6],
$$ u^\mu=\dot x^\mu;~~~\sigma^{\mu\nu}=\Lambda_\lambda^{\quad\mu}\dot\Lambda
^{\lambda\nu};~~~\Lambda_\lambda^{\quad \mu}\Lambda^{\lambda\nu}
=\Lambda^\mu_{\quad \lambda}\Lambda^{\nu\lambda}=g^{\mu\nu}$$
$$ g^{00}=-g^{11}=-g^{22}=1.$$
A specific (2+1-) dimensional feature of this
model should be pointed out. It was shown in [7], that in 3+1-dimensions,
the simple coupling term $\sigma .F$ was unphysical since it puts undue
restrictions on the external field, because of the constraint structure and
more complicated couplings had to be introduced. One can recognise the
phenomenological electric and magnetic dipole moment terms [11] in the
$D$-term,
but constructed here out of the basic degrees of freedom.

The conjugate momenta are directly obtained as
\be P^\mu = -{{\partial L}\over{\partial u_\mu}} = -{1\over 2}({\cal
L})^{-1\over 2}
(2M^2u^\mu + MJ\epsilon ^{\mu\nu\lambda}\sigma_{\nu\lambda}) + eA^\mu ,\ee
\be S^{\mu\nu}=-{{\partial L}\over{\partial \sigma_{\mu\nu}}} = -({\cal
L})^{-1\over 2}
(J^2\sigma^{\mu\nu} + MJ\epsilon^{\mu\nu\lambda}u_\lambda) + DF^{\mu\nu}. \ee
Defining $\Pi^\mu =P^\mu-eA^\mu$ and $\Sigma ^{\mu\nu}=S^{\mu\nu}-DF^{\mu\nu}$,
the primary (but not independent) constraints are
\be \Pi^2=M^2;~~~ {1\over
2}\epsilon^{\mu\nu\lambda}\Sigma_{\mu\nu}\Pi_\lambda=MJ;~~~
\Sigma^2=2J^2, \ee
\be V^\mu=\Sigma^{\mu\nu}\Pi_\nu=0. \ee
In the covariant framework, the natural choice for the independent First Class
Constraints (FCC) are the mass shell condition and the Pauli-Lubanski scalar,
the first two relations of (4) respectively.
 This is the starting point of the Jackiw-Nair construction [2]. However,
in the fixed time or Hamiltonian scheme, more useful is the choice of the
mass shell condition and (5), used here, which
 was advocated in [7]. Note that actually (5) consists of {\it two}
independent relations since $\Pi_\mu V^\mu=0$ and this Second Class (SC) pair
together with mass shell condition implies the Pauli-Lubanski relation.
 A gauge fixing for the latter in the former alternative
makes the number of constraints same in both the cases. (This point was
not stressed in [6].) However we are not finished with the constraints yet.
The so called Weyssenhoff condition (5) ensures that in the particle
rest frame, the spin has a single component $S^{12}$. Consequently one must
introduce additional constraints that restrict the number of angular degrees
of freedom to one in the rest frame as well. This is achieved by the constraint
\be \chi^\mu=\Lambda^{0\mu}-{{\Pi^\mu}\over M} \approx 0. \ee
 The identity $\Pi^\mu \chi_\mu
=-{1\over 2}M\chi ^2$ shows that only two $\chi$'s are independent.

The fundamental Poisson Brackets (PB) are the following:
\be \{\Pi^\mu,x^\nu\}=-g^{\mu\nu};~~~ \{\Pi^\mu,\Pi^\nu\}=-eF^{\mu\nu} \ee
\be \{S^{\mu\nu},S^{\alpha\beta}\}=S^{\mu\alpha}g^{\nu\beta}-S^{\mu\beta}
g^{\nu\alpha}+S^{\nu\beta}g^{\mu\alpha}-S^{\nu\alpha}g^{\mu\beta}, \ee
with all other PB's vanishing. Note that since $\{\Pi^\mu,\Pi^\nu\}$ PB is
nontrivial,
the mass shell condition in (4) has to be modified to
\be \Pi^2-M^2+{{2eF_{\mu\alpha}\Pi^\mu}\over{\Pi^2-{e\over 2}F.\Sigma}}V^\alpha
\ee
to make it an FCC. But this extra term can be ignored since we immediately move
on to the Dirac Brackets (DB) generated by $V^\mu$ and hence can use $V^\mu =0$
strongly. Let us now comput the Dirac Brackets (DB) for the SCC's $V^i$ and
$\chi^j$ for $I=1,2$ since only two each of the $V^\mu$ and $\chi^\mu$ are
independent.
This manifestly non-covariant algebra can be avoided by
an elegant trick, (for details see [7]) and equivalently one can construct the
DB's
for $V^\mu$ and $\chi^\nu$. After inverting the constraint matrix consisting
of PB's of the SCC's, we get the first stage $(DB)_I$ of any two generic
operators as
$$ \{A,B\}_I=\{A,B\}_{PB}+\{A,V^\mu \}({{eF_{\mu\nu}}\over{m^2 M^2}})\{V^\nu
,B\}-\{A,\chi^\mu\}(S_{\mu\nu} $$
\be +{D\over M^2}(\Pi_\mu F_{\nu\lambda}-\Pi_\nu F_{\mu
\lambda})\Pi^\lambda)\{\chi^\nu,B\}+\{A,V^\mu\}{1\over M}\{\chi_\mu,B\}
-\{A,\chi^\mu\}{1\over M}\{V_\mu,B\} \ee
where $m^2=M^2-{e\over 2}S.F$. Thus the $O(F)$ DB's relevant to our present
interest are
$$ {\{\Pi^\mu,\Pi^\nu\}}_I=-eF^{\mu\nu};~~~ {\{\Pi^\mu,x^\nu\}}_I=-g^{\mu\nu}
-{e\over M^2}F^{\mu\alpha}S^\nu_{\quad\alpha}~ ; $$
\be {\{\Pi^\mu,S^{\nu\lambda}\}}_I={e\over M^2}F^\mu_{\quad\alpha}(\Pi^\nu
S^{\alpha
\lambda}-\Pi^\lambda S^{\alpha\nu}). \ee
Note that to $O(F)$, the $D$ terms do not appear.
We are still left with the FCC $\Pi^2-M^2$. This reflects the arbitrariness
present in the definition of the parameter $\tau$ in $u^\mu={dx^\mu\over
d\tau}$.
We gauge fix $\tau$ to be the proper time, $x^0-\tau\approx 0$ and now compute
the final set of DB's for the pair of Scc's $\Pi^2-M^2\approx0$ and
$x^0-\tau\approx 0$. The final brackets are
\be \{\Pi^\mu,S_\gamma\}^*={e\over M^2}\epsilon_{\gamma\nu\lambda}F^\mu
_{\quad\alpha}
\Pi^\nu S^{\alpha\lambda} +{{e\epsilon_{\gamma\nu\lambda}\Pi_\alpha F^{\mu
\alpha}}\over {\Pi^0 m^2}}S^{0\nu}\Pi^\lambda,\ee
\be \{\Pi^\mu,\Pi^\nu\}^*=-eF^{\mu\nu}+{{e\Pi_\alpha}\over \Pi^0}(F^{\mu\alpha}
g^{0\nu}+F^{\alpha\nu}g^{\mu o}) \ee
where $ S_\gamma={1\over 2}\epsilon_{\gamma\mu\nu}S^{\mu\nu}$ is the
relativistic generalization of the spin. Now we are equipped to derive the
Hamiltonian equations of motion.

The canonical Hamiltonian $H$ vanishes, (the
theory being invariant under redefinitions of $\tau$), but there are extra
contributions to $H$ since the gauge fixing $x^0-\tau\approx 0$ is time
dependent.
 This extra term is obtained in the standard way [12] and we end up with a
 non zero $H$,
 \be H=\Pi^0 ={(M^2-\Pi^i\Pi_i)}^{1\over 2}. \ee
 Hence the equations of motion are
 \be {\dot \Pi}^\mu=\{H,\Pi^\mu\}^*={{e\Pi_i}\over\sqrt {(M^2-\Pi^i\Pi_i)}}
 (F^{i\mu}-F^{i0}g^{0\mu})\ee
 \be {\dot S}^\mu=\{H,S^\mu\}^*=-{e\over {M^3}}\epsilon^{\mu\nu\lambda}\Pi_i
F^{ij}\Pi_\nu
 S_{j\lambda}. \ee
 One can at once check that
 \be {\dot \Pi}^0=0;~~~ \dot S^0=0;~~~{\dot \Pi}^i=-{e\over M}F^{ij}\Pi_j;
 ~~~\dot S^i=-{e\over M^3}(\Pi.S)F^{ij}\Pi_j. \ee
 Notice that $\Pi.S=MJ$ from (4) and so with the consistent choice ${J\over M}
 \Pi^\alpha=S^\alpha$, we see that the two equations obtained separately are
 infact identical. We emphasize that this proof of the fact that (generalized)
momentum and spin are
 parallel for anyons is rigorous since here the relation ${J\over M}\Pi^\alpha
 =S^\alpha$ is an operator identity, (valid even after quantisation since we
 have used the Dirac Brackets), whereas stating this relation only from (4)
would
 be purely classical. This is equivalent to the identification made in [8].
{}From
 now on we will consider only the $S^\alpha$ equation, which is the appropriate
 BMT equation.
 Understandably the equations are not manifestly covariant. This is simply
 because we are working in the fixed-time or Hamiltonian framework. Indeed one
 can see that we are in the particle rest frame, (since we fixed $\tau$ to be
the proper time).
 Hence the BMT equation (17) is at once generalized to
 \be \dot S^\alpha=-{e\over M}F^{\alpha\beta}S_\beta.\ee
 Comparing this with the original BMT equation [8] we verify that the
gyromagnatic
 ratio for anyons is 2.

We conclude with two comments regarding the external field formulation as well
as
some future directions of study. Firstly remember that we have always assumed
the fields to be external, meaning that they are c-number space-time functions,
devoid of a nontrivial PB structure. However, the nonzero $\{x^\mu,x^\nu\}$
DB's , originating from (10), (and present even without interactions),
introduce a nonvanishing $\{A^\mu,A^\nu\}$. Does this mean that the $A^\mu$'s
are no longer external? According to Hanson and Regge [7], one should include
the gauge kinetic term as well from the beginning and then proceed. This will
make the gauge fields dynamical variables from the start.
Secondly, since the DB structure is drastically altered from the PB's,
specially with respect to the $\{x^\mu,x^\nu\}$ algebra, canonical
quantization is problematic. Hence one can either use directly the
Wigner-Pryce [7] variables or one can try to solve the constraints
perturbatively, as was done in [8].

The future prospects of this kind of Lagrangian and Hamiltonian
formulations look quite promising. Since there is an explicit Lagrangian to
start from, one can try to construct models for two interacting anyons. Also
gravitational interactions can be introduced in a straightforward manner.
Work is in progress along these directions.


\begin{thebibliography}{99}

\bibitem 1 F.Wilczek, {\it Fractional Statistics and Anyon Superconductivity}
(world Scientific, 1990)
\bibitem 2 R.Jackiw and V.P.Nair, Phys.Rev.{\bf D 43} 1933(1991)
\bibitem 3 M.Chaichian, R.Gonzalez Felipe and D.Lois Martinez, Phys.Rev.Lett.
{\bf 71} 3405(1993)
\bibitem 4 J.-H.Cho, S.Hyun and H.-J.Lee, Phys.Lett.{\bf B 327} 274(1994);
J.-H.Cho and J.-K.Kim, Phys.Lett.{\bf B 332} 118(1994)
\bibitem 5 R.Jackiw and V.P.Nair, Comment on Phys.Rev.Lett. {\bf 71}
3405(1993), Preprint 1994;
N.Banerjee, R.Banerjee and S.Ghosh, Relativistic theory of free anyon
revisited,
Preprint SINP/TNP/94-08
\bibitem 6 S. Ghosh, Spinning particles in 2+1-dimensions, to appear in
Phys.Lett. {\bf B} 1994
\bibitem 7 A.J.Hanson and T.Regge, Ann.Phys. (N.Y.), {\bf 87} 498(1974)
\bibitem 8 C.Chou, V.P.Nair and A.P.Polychronakos, Phys.Lett.{\bf B 304}
105(1993)
\bibitem 9 V.Bargmann,L.Michel and V.L.Telegdi, Phys.Rev.Lett. {\bf 2}
435(1959)
\bibitem {10} P.A.M.Dirac, {\it Lectures on Quantum Mechanics} (Yeshiva
University Press, 1964)
\bibitem {11} A.O.Barut, {\it Electrodynamics of the Classical Theory of
Fields and Particles}, (Macmillan Co.,N.Y. 1964)
\bibitem {12} D.M.Gitman and I.V.Tyutin, {\it Quantization of Fields with
Constraints}, (Springer-Verlag)

\end{thebibliography}
\end{document}